\documentclass{osa-article-arXiv}

\journal{oe}

\usepackage{graphicx}
\usepackage{dcolumn}
\usepackage{bm}
\usepackage{hyperref}
\usepackage[mathlines]{lineno}
\usepackage{amsmath}
\usepackage{esint}
\usepackage{siunitx}
\usepackage[ruled]{algorithm2e}
\SetKwComment{Comment}{/* }{ */}
\usepackage[capitalise]{cleveref} 

\begin{document}

\title{Towards optimal point spread function design for resolving closely spaced emitters in three dimensions}

\author{James M.\ Jusuf\authormark{1,4} and Matthew D.\ Lew\authormark{1,2,3,*}}

\address{
\authormark{1}Department of Electrical and Systems Engineering, Washington University in St. Louis, MO 63130, USA\\
\authormark{2}Center for the Science and Engineering of Living Systems, Washington University in St. Louis, MO 63130, USA \\
\authormark{3}Institute of Materials Science and Engineering, Washington University in St. Louis, MO 63130, USA\\
\authormark{4}Present address: Department of Biological Engineering, Massachusetts Institute of Technology, Cambridge, MA 02139, USA}

\email{\authormark{*}mdlew@wustl.edu} 

\homepage{https://lewlab.wustl.edu}


\begin{abstract}
The past decade has brought many innovations in optical design for 3D super-resolution imaging of point-like emitters, but these methods often focus on single-emitter localization precision as a performance metric. Here, we propose a simple heuristic for designing a point spread function (PSF) that allows for precise measurement of the distance between two emitters. We discover that there are two types of PSFs that achieve high performance for resolving emitters in 3D, as quantified by the Cram\'{e}r-Rao bounds for estimating the separation between two closely spaced emitters. One PSF is very similar to the existing Tetrapod PSFs; the other is a rotating single-spot PSF, which we call the crescent PSF. The latter exhibits excellent performance for localizing single emitters throughout a 1-\textmu{}m focal volume (localization precisions of 7.3~nm in $x$, 7.7~nm in $y$, and 18.3~nm in $z$ using 1000 detected photons), and it distinguishes between one and two closely spaced emitters with superior accuracy ($25$-$53\%$ lower error rates than the best-performing Tetrapod PSF, averaged throughout a 1-\textmu{}m focal volume). Our study provides additional insights into optimal strategies for encoding 3D spatial information into optical PSFs.
\end{abstract}

\section{\label{sec:intro}Introduction}
Recent years have brought significant advances in fluorescence nanoscopy, with three-dimensional (3D) single-molecule tracking and super-resolution microscopy \cite{VonDiezmann2017,Lelek2021} approaching atomic resolution. A cornerstone of these methods and of the Nobel Prize in Chemistry 2014 \cite{Betzig2015,Hell2015,Moerner2015} is the switching of a molecule's emissive state, i.e., experimenter-controlled ``blinking,'' which enables individual molecules to be localized independently by minimizing the overlap between their individual images, or point spread functions (PSFs). Localizations gathered over time are stitched together to construct a final super-resolved image. Combined with single-molecule blinking, imaging methods such as interferometry \cite{Shtengel2009,Aquino2011,Huang2016,Backlund2018,Bates2022}, fluorescence lifetime imaging near metallic or carbon surfaces \cite{Isbaner2018,Ghosh2019,Thiele2022}, and structured illumination with active feedback \cite{Gwosch2020} have all been demonstrated to localize single fluorescent molecules in 3D space with precisions approaching 0.1-1~nm.

However, because the emissive state of any particular fluorophore can only be controlled probabilistically, 3D nanoscopes routinely must detect, resolve, and estimate the positions of molecules whose PSFs overlap. Methods to improve the resolvability of pairs of emitters laterally separated along the $x$- and $y$-axes include PSF engineering for direct imaging \cite{Paur2018}, as well as the use of finite optical structures, such as waveguides, for separating spatial modes in indirect imaging \cite{Tsang2016}. Additionally, neural networks have achieved impressive performance for the joint task of designing a PSF and resolving dense constellations of emitters within noisy images with high accuracy and resolution \cite{Nehme2020, Nehme2021}. Recent studies have also examined the fundamental performance limits of localizing emitter pairs in 3D using quantum estimation theory \cite{Yu2018, Prasad2019}.

Despite decades of innovation, several outstanding questions remain. First, how can we express the joint task of resolving and localizing overlapping emitters in 3D mathematically as a performance metric or cost function? Recent advances by Yoav Shechtman and colleagues \cite{Nehme2020, Nehme2021} elegantly adapt a similarity statistic, the Jaccard index, to evaluate if a neural network accurately identifies and localizes single molecules within test images. However, this strategy requires careful design and generation of test data that accurately models the imaging task at hand. Secondly, given a suitable metric, is there a globally optimal PSF that achieves the best possible performance? Or, are there a few or perhaps many designs that all perform similarly? Finally, are there general design principles that we may interpret as optimal for resolving emitters in 3D? Often, PSF designs resulting from numerical optimization studies are difficult to interpret and generalize. Examples of PSFs that are widely used for 3D single-molecule localization include both rotating PSFs (e.g., the double-helix \cite{Pavani2009}) and expanding/translating PSFs (e.g., the Tetrapod \cite{Shechtman2015} and Airy \cite{Jia2014,Zhou2020} families). Addressing these questions will have numerous implications for the advancement of fluorescence nanoscopy. Imaging dense emitter configurations in 3D is key in many biophysical applications, including studies of subcellular localization of mRNA molecules \cite{Eng2019}, chromatin looping dynamics \cite{Brandao2020}, and protein organization within cilia \cite{Bennett2020}. From a methods perspective, the ability to detect and resolve closely spaced emitters in 3D is critical for high-performance 3D super-resolution imaging \cite{Sage2019}.

In this theoretical study, we address the questions above by engineering PSFs for resolving pairs of emitters in all three dimensions. We propose a cost function that quantifies a PSF's performance based on the precision to which it can be used to measure the distance between two emitters with small separation in both the lateral and axial directions. We then apply a gradient-descent algorithm to search for PSF(s) that minimize this cost function. Surprisingly, all runs of the algorithm converge to one of two designs, regardless of the initial condition. One is a ``single-spot'' PSF that rotates as a function of emitter depth, which we call the crescent PSF, and the other is a ``double-spot'' PSF that mimics existing Tetrapod PSFs by expanding laterally as a function of emitter depth. We quantify the theoretical performance of these PSFs by calculating the classical Cram\'er-Rao bounds (CRBs) for estimating positional quantities in one- and two-emitter configurations. Moreover, a likelihood-ratio test on simulated data demonstrates that compared to other engineered PSFs, the crescent PSF allows for distinguishing between one emitter versus two axially separated emitters with superior accuracy.

\section{Designing imaging systems for resolving closely spaced emitter pairs in three dimensions}

\subsection{Mathematical framework}
\begin{figure}[t!]
\centering\includegraphics[width=5.25in]{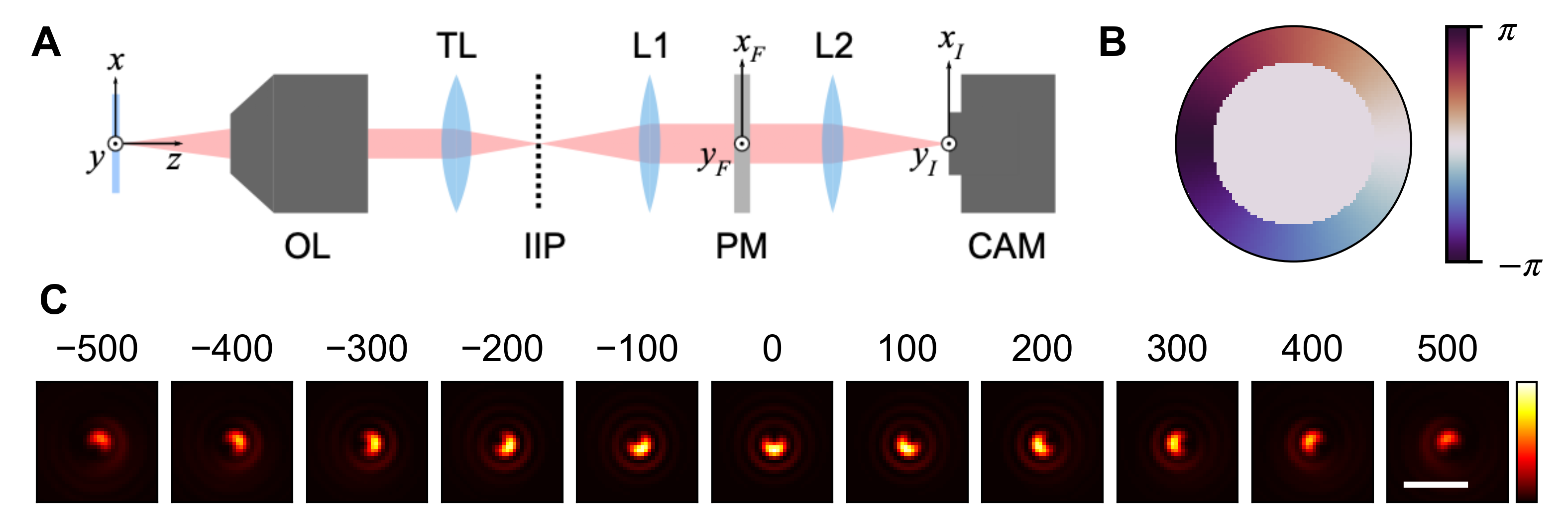}
\caption{Optical system for implementing the crescent PSF. (A)~A 4f system comprising two lenses (L1, L2) and a phase mask (PM) is attached to a standard microscope with an objective lens (OL) and tube lens (TL). The intermediate image plane (IIP) and camera (CAM) image planes are conjugate to the focal plane of OL. The coordinates $x_F$ and $y_F$ are scaled such that the aperture has radius $\mathrm{NA}/n$. (B)~The optimized PM that produces the crescent PSF. The colorbar represents phase in radians. (C)~Images of the optimized crescent PSF as a function of emitter depth in nm. The color scale is in arbitrary units of intensity. Scale bar = 1~\textmu\text{m}.}
\label{fig:Schematic}
\end{figure}

We frame our study on resolving emitters in 3D as a PSF design problem. Formally, the PSF of an optical system is defined as the image produced by a single idealized point emitter. In this paper, we restrict ourselves to modeling such emitters, taking them to represent single fluorescent molecules. To express the PSF of a microscope mathematically, we begin by considering an emitter located at $(x_0, y_0, z_0)$ in object space, which produces the following classical wave function in the Fourier plane of the microscope \cite{Petrov2017}:
\begin{equation}\label{eq:psi}
    \psi(x_F, y_F; x_0, y_0, z_0) = A\left(1-r_F^2\right)^{-1/4}\text{circ}\left(\frac{nr_F}{\mathrm{NA}}\right) \exp\left[\frac{i2\pi n}{\lambda_0}\left(x_0 x_F + y_0 y_F + z_0\sqrt{1-r_F^2}\right)\right].
\end{equation}
Here, $x_F$ and $y_F$ are the spatial coordinates in the Fourier plane, $n$ is the index of refraction of the medium surrounding the emitter, $\mathrm{NA}$ is the numerical aperture of the microscope, $\lambda_0$ is the free-space wavelength of the emitter, and $r_F = \sqrt{x_F^2+y_F^2}$. The circular aperture function $\text{circ}$ denotes the disk-shaped support of the wavefunction in Fourier space and is defined as
\begin{equation}
    \text{circ}(\rho) = \begin{cases} 
      1 & \rho \leq 1 \\
      0 & \text{otherwise,}
   \end{cases}
\end{equation}
and the normalization factor $A$ is given by
\begin{equation}
    A = \left[2\pi \left(1-\sqrt{1-(\mathrm{NA}/n)^2}\right)\right]^{-1/2},
\end{equation}
ensuring that $\iint dx_F\, dy_F\, |\psi(x_F, y_F; x_0, y_0, z_0)|^2 = 1$. We also consider a phase mask (PM) in the Fourier plane, described by a function $\varphi_\text{mask}(x_F, y_F)$ with range $[-\pi,\pi)$, which enables us to design the image of the emitter with high photon efficiency (\cref{fig:Schematic}A). The final PSF is given by
\begin{equation}\label{eq:psf}
    I(x_I, y_I; x_0, y_0, z_0) = \lvert \mathcal{F}\{\psi(x_F, y_F;x_0, y_0, z_0)\exp\left[i\varphi_\text{mask}(x_F, y_F)\right]\}\rvert^2,
\end{equation}
where $\mathcal{F}$ is a two-dimensional Fourier transform.

In single-molecule imaging, the precision to which emitters can be localized is limited by the probabilistic nature of photon detection. We borrow a powerful result from estimation theory called the Cram\'er-Rao bound (CRB): given a quantity of interest $\theta$, which could be an emitter's $x$-coordinate, for instance, the CRB provides a lower bound on the variance of any unbiased estimator $\hat{\theta}$ of $\theta$:
\begin{equation}\label{eq:crb}
    \text{Var}(\hat{\theta}) \geq \left(\sigma^\text{(CRB)}_\theta\right)^2.
\end{equation}
Therefore, $\sigma^\text{(CRB)}_\theta$ can be used to quantify the precision with which a given optical system can estimate $\theta$ from a noisy image. It is related to the Fisher information $F(\theta)$ by \cite{Chao2016}
\begin{equation}\label{eq:crb_fi}
    \left(\sigma^\text{(CRB)}_\theta\right)^2 = [F(\theta)]^{-1}.
\end{equation}
Modeling photon detection as a Poisson process, for an image formed by $N_\text{sig}$ signal photons from the emitter and a constant background flux of $N_\text{bg}$~photons per pixel, the Fisher information with respect to $\theta$ is given by
\begin{equation}\label{eq:fisher-information}
    F(\theta) = \iint dx_I\, dy_I\, \frac{N_\text{sig}^2}{N_\text{sig} p(x_I, y_I|\theta)+N_\text{bg}}\left[\frac{\partial}{\partial \theta}p(x_I, y_I|\theta)\right]^2,
\end{equation}
where $p(x_I, y_I|\theta)$ represents the probability density of photon detection over image space. For a single point emitter, this is equivalent to the PSF, given in \cref{eq:psf}:
\begin{equation}\label{eq:prob-single-emitter}
    p(x_I, y_I|x_0, y_0, z_0) = I(x_I, y_I;x_0, y_0, z_0).
\end{equation}

We extend the formalism above to calculate CRBs in the case of two emitters. For two equally bright incoherent point emitters located at $(x_c \pm \Delta x/2, y_c, z_c)$, where the subscript $c$ is used to denote their ``centroid'' coordinates, the probability density becomes the average of the contributions from the two emitters:
\begin{equation}\label{eq:prob-two-emitters}
    p(x_I, y_I|x_c, \Delta x, y_c, z_c) = \frac{1}{2}\left[I\left(x_I,y_I;x_c + \frac{\Delta x}{2}, y_c, z_c\right) + I\left(x_I,y_I;x_c - \frac{\Delta x}{2}, y_c, z_c\right)\right].
\end{equation}
If $\Delta x$ is sufficiently small, then $I(x_I, y_I; x_c \pm \Delta x/2, y_c, z_c)$ can be expanded as a Taylor series along the direction of separation, which gives
\begin{equation}\label{eq:taylor_x}
    p(x_I, y_I|x_c, \Delta x, y_c, z_c) = I(x_I, y_I; x_c, y_c, z_c) + \frac{\Delta x^2}{8} \frac{\partial^2}{\partial x^2}I(x_I, y_I; x, y_c, z_c)|_{x=x_c} + \dots.
\end{equation}
By ignoring all higher order terms and substituting \cref{eq:taylor_x} into \cref{eq:fisher-information}, we obtain approximations for the Fisher information $F(\Delta x)$ and corresponding precision $\sigma_{\Delta x}^\text{(CRB)}$ for estimating $\Delta x$. For convenience, we define
\begin{equation}\label{eq:sx}
    S_{x} \equiv \iint dx_I\, dy_I\, \frac{N_\text{sig}^2}{N_\text{sig} I(x_I, y_I;x_c, y_c, z_c)+N_\text{bg}} \left[\frac{\partial^2}{\partial x^2}I(x_I, y_I; x, y_c, z_c)|_{x=x_c}\right]^2.
\end{equation}
(Note that $N_\text{sig}$ now refers to the total number of photons from both emitters combined.) The Fisher information can be approximated as $F(\Delta x) \approx \Delta x^2 S_{x}/16$ and the estimation precision as
\begin{equation}\label{eq:sigma_delta_x}
    \sigma_{\Delta x}^\text{(CRB)} \approx \frac{4}{\Delta x\sqrt{S_x}}.
\end{equation}
We repeat this process for emitter pairs with $y$- and $z$-axis separation. For two emitters located at $(x_c, y_c\pm\Delta y, z_c)$, we define
\begin{equation}\label{eq:sy}
    S_{y} \equiv \iint dx_I\, dy_I\, \frac{N_\text{sig}^2}{N_\text{sig} I(x_I, y_I;x_c, y_c, z_c)+N_\text{bg}} \left[\frac{\partial^2}{\partial y^2}I(x_I, y_I; x_c, y, z_c)|_{y=y_c}\right]^2,
\end{equation}
and the estimation precision for $\Delta y$ is approximately
\begin{equation}\label{eq:sigma_delta_y}
    \sigma_{\Delta y}^\text{(CRB)} \approx \frac{4}{\Delta y\sqrt{S_y}}.
\end{equation}
Similarly, for two emitters located at $(x_c, y_c, z_c\pm\Delta z)$, we define
\begin{equation}\label{eq:sz}
    S_{z} \equiv \iint dx_I\, dy_I\, \frac{N_\text{sig}^2}{N_\text{sig} I(x_I, y_I;x_c, y_c, z_c)+N_\text{bg}} \left[\frac{\partial^2}{\partial z^2}I(x_I, y_I; x_c, y_c, z)|_{z=z_c}\right]^2,
\end{equation}
and the estimation precision for $\Delta z$ is approximately
\begin{equation}\label{eq:sigma_delta_z}
    \sigma_{\Delta z}^\text{(CRB)} \approx \frac{4}{\Delta z\sqrt{S_z}}.
\end{equation}
A comparison of the approximations in \cref{eq:sigma_delta_x,eq:sigma_delta_y,eq:sigma_delta_z} against their true values is shown in Appendix \ref{subsec:CRBapprox}.

\subsection{Point-spread function optimization}\label{subsec:optimization}
In order to engineer a PSF that excels at resolving closely spaced pairs of emitters in 3D, we wish to find a phase mask that simultaneously minimizes $\sigma^\text{(CRB)}_{\Delta x}$, $\sigma^\text{(CRB)}_{\Delta y}$, and $\sigma^\text{(CRB)}_{\Delta z}$ for two equally bright incoherent point emitters in various spatial configurations. The following cost function simultaneously accounts for all three quantities:
\begin{equation}
    C_0(\varphi_\text{mask}) = \left(\sigma_{\Delta x}^\text{(CRB)}\right)^2 + \left(\sigma_{\Delta y}^\text{(CRB)}\right)^2 + \left(\sigma_{\Delta z}^\text{(CRB)}\right)^2.
\end{equation}
The right-hand side implicitly depends on $\varphi_\text{mask}$ according to the equations in the previous subsection. It is worth noting that $\sigma^\text{(CRB)}_{\Delta x}$ is well-defined for any placement of two emitters, including one with nonzero $y$- and $z$- separation. However, to reduce the computational complexity of the problem, we calculate $\sigma^\text{(CRB)}_{\Delta x}$ for emitters separated along the $x$-axis only (as in \cref{eq:sx,eq:sigma_delta_x}). Similarly, we calculate $\sigma^\text{(CRB)}_{\Delta y}$ for emitters with $y$-separation only (as in \cref{eq:sy,eq:sigma_delta_y}) and $\sigma^\text{(CRB)}_{\Delta z}$ for emitters with $z$-separation only (as in \cref{eq:sz,eq:sigma_delta_z}). This convention is used throughout the remainder of the paper.

Besides $\varphi_\text{mask}$, $C_0$ also depends on a multitude of variables, namely $x_c$, $y_c$, $z_c$, $\Delta x$, $\Delta y$, $\Delta z$, $N_\text{sig}$, and $N_\text{bg}$, and the engineered PSF should minimize $C_0$ over a reasonable domain of these parameters. 
The optical system used in this paper is assumed to be shift-invariant, and since the integrals in Eqs. (11), (13) and (15) are evaluated over the entire image space, all $\sigma^\text{(CRB)}$ are independent of $x_c$ and $y_c$. We therefore fix $x_c=y_c=0$ without loss of generality. We also set $N_\text{sig}=1000$ and $N_\text{bg}=10$, which are typical for single molecule-imaging experiments. Furthermore, our approximate expression for $\sigma_{\Delta x}^\text{(CRB)}$ (\cref{eq:sigma_delta_x}) consists of two separate factors, one that only depends on the PM and one that only depends on $\Delta x$. This implies that a PM that performs well at one value of $\Delta x$ will perform well at all other sufficiently small values of $\Delta x$, so there is no need to include $\Delta x$ in the cost function. The same argument applies for $\sigma_{\Delta y}^\text{(CRB)}$ and $\sigma_{\Delta z}^\text{(CRB)}$ (\cref{eq:sigma_delta_y,eq:sigma_delta_z}). It therefore suffices to minimize
\begin{equation}
    C_1(\varphi_\text{mask}) = \frac{1}{S_x} + \frac{1}{S_y} + \frac{1}{S_z},
\end{equation}
which now only depends on the PM and $z_c$.

To find a PM that minimizes $C_1$ over a range of possible $z_c$ values, we define $C_2$ to be proportional to the mean value of $C_1$ over 15 equally-spaced values of $z_c$  between $-500$~nm and $500$~nm:
\begin{equation}\label{eq:C2}
    C_2(\varphi_\text{mask}) = \frac{\beta}{15} \times \sum_{z_c \in \{-500~\text{nm}, \dots, 500~\text{nm}\}} C_1(\varphi_\text{mask};z_c).
\end{equation}
The constant $\beta$ can be chosen freely to scale the values of $C_2$ to any desired order of magnitude.
 
We apply a gradient descent optimization algorithm to find PMs that minimize $C_2$. To carry out the computations, the continuous function $\varphi_\text{mask}$ is parametrized as a real-valued $256\times256$ matrix whose elements represent a discretely-sampled (i.e., pixelated) PM. A pseudocode version of the gradient descent algorithm is shown in \cref{alg:grad_descent}. We use the Python programming language (version 3.7.11) and TensorFlow machine learning library (version 2.0.0) to implement the algorithm. While the optimization task is not a machine learning problem, we choose TensorFlow for its automatic differentiation capabilities, specifically the \texttt{GradientTape} function. All PSFs are calculated for an imaging system with $\mathrm{NA}=1.4$, $n=1.518$, $\lambda_0=550~\text{nm}$, magnification $M=111.11$, and 4f lens focal length $f=150$~mm. For a PM that is $N=256$ pixels wide with a pixel size of $d_\text{PM}=\SI{49.58}{\micro\meter}$, the image pixel size is
\begin{equation}
    d_I = \frac{\lambda f}{d_\text{PM}MN} = \SI{58.5}{nm}
\end{equation}
when calculating the PSF using a 2D discrete Fourier transform. These values are chosen to reflect a typical laboratory setup and are used to generate all the results in this paper. The nonzero values of the PM matrix are restricted to a centered disk of radius $38.12~\text{px}$, which represents the aperture of the microscope. We also set $\beta = 1.6\times10^{29}$ in \cref{eq:C2} such that when distances are inputted in meters, typical values of $C_2$ are on the order of 10.

\begin{figure}[ht!]
\centering
\begin{minipage}{0.95\linewidth}
    \begin{algorithm}[H]
        \SetAlgoLined
        \caption{Gradient descent for designing PM $\varphi_{\text{mask}}$ to minimize cost function $C_2$ (\cref{eq:C2}) \label{alg:grad_descent}}
        $i \gets 0$ \Comment*[r]{$i_{\text{max}}$ sets the number of iterations}
        \While{$i < i_\text{max}$} 
        {
            $\bm{v} \leftarrow \bm{v} - \alpha\nabla_{\bm{v}} C_2(\varphi_{\text{mask},\bm{v}})$ \Comment*[r]{$\bm{v}$ is a vector of PM pixel values, $\alpha$ is a user-defined learning rate, $\varphi_{\text{mask},\bm{v}}$ is the PM corresponding to $\bm{v}$}
            $i \leftarrow i + 1$\;
        }
        \Return $\bm{v}$\;
    \end{algorithm}
\end{minipage}
\end{figure}

\begin{figure}[hbt!]
\centering\includegraphics[width=5.25in]{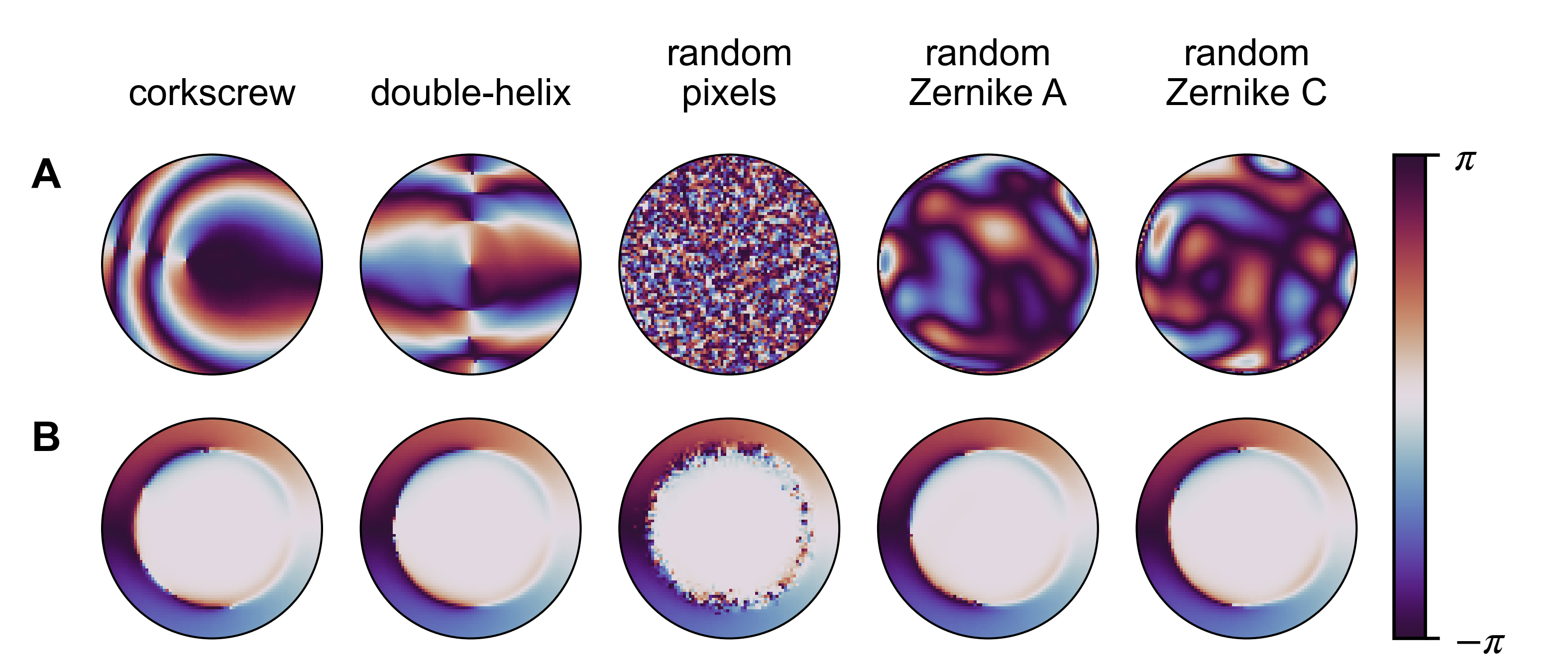}
\caption{Optimality and stability of the crescent PSF. (A)~The various initial PMs, including those of the corkscrew \cite{Lew2011} and double-helix \cite{Pavani2009} PSFs, that converge to the crescent PM when optimized by \cref{alg:grad_descent}. (B)~The final PMs, standardized as described in Appendix \ref{subsec:normalizingPMs}. The PSFs generated by the final PMs are virtually identical to the PSF in \cref{fig:Schematic}C. The colorbar represents phase in radians.}
\label{fig:crescentPMs}
\end{figure}

Since there is no reason to assume that $C_2$ is a convex function of the PM's pixel values, the local minimum to which gradient descent converges may depend on the initial condition. Thus, we run the algorithm from 14 different initial PMs (Fig.~\ref{fig:crescentPMs}A and Appendix \ref{subsec:tetraLikePMs}). Seven of the initial PMs correspond to existing PSFs engineered for 3D localization of single emitters: these are the corkscrew \cite{Lew2011} and double-helix \cite{Pavani2009} PSFs, two PSFs from the Tetrapod family \cite{Shechtman2015} (termed tetra2 and tetra3 in this paper), two astigmatic PSFs \cite{Kao1994}, and one PSF from the twin Airy family \cite{Zhou2020}. Another PM consists of pixels with random values, i.e., each pixel value is independently sampled uniformly between $-\pi$ and $\pi$. The six remaining PMs are generated as linear combinations of 63 Zernike polynomials on the circular domain of the PM with $2\leq n \leq 10$, with coefficients independently sampled from a normal distribution with a mean of $0$ and standard deviation of $0.05$. The first three Zernike polynomials ($n=0,1$) are excluded since they do not affect $C_2$. The learning rate $\alpha$ is set to 25 for the corkscrew, double-helix, tetra2, tetra3, astigmatism, and twin Airy initial conditions, 1 for the random pixels initial condition, and 10 for the random Zernike initial conditions. The algorithm converges in all cases (Appendix \ref{subsec:costFxnDescent}).

We standardize the final PMs resulting from gradient descent such that their PSFs are centered laterally in the image plane and similar PSFs share the same orientation (Appendix \ref{subsec:normalizingPMs}). It is then evident that the PMs fall into two categories: (1)~Five PMs consist of a ring-shaped ramp around a circular region of constant zero phase and are shown in \cref{fig:crescentPMs}B. The results are essentially identical except for minor imperfections at the boundary between the two regions. We therefore design an idealized version of the PM where the boundary is well-defined (Appendix \ref{subsec:idealizedCrescentPM}); the final PM is shown in \cref{fig:Schematic}B. This PM produces a single-spot PSF that rotates as a function of emitter depth (\cref{fig:Schematic}C), which we call the crescent PSF. Its rotation and shape bear a striking resemblance to the corkscrew PSF \cite{Lew2011, Prasad2013}, even though its PM is remarkably different. (2)~The other nine final PMs from gradient descent produce a PSF that resembles those of the existing Tetrapod family (Appendix \ref{subsec:tetraLikePMs}).

\section{Precision of crescent PSF for one- and two-emitter localization}

To compare the performance of the crescent PSF against that of existing engineered PSFs, we calculate $\sigma^\text{(CRB)}$ for various positional quantities within a 1-\textmu{}m depth range ($-500~\text{nm}\leq z_c \leq 500~\text{nm}$) for one- and two-emitter configurations. All quantities are calculated using the same microscope parameters as in gradient descent. For the existing engineered PSFs, we consider the double-helix as well as tetra2, which is the Tetrapod PSF with the best performance within our chosen range of $z_c$.

\begin{figure}[htb!]
\centering\includegraphics[width=4.5in]{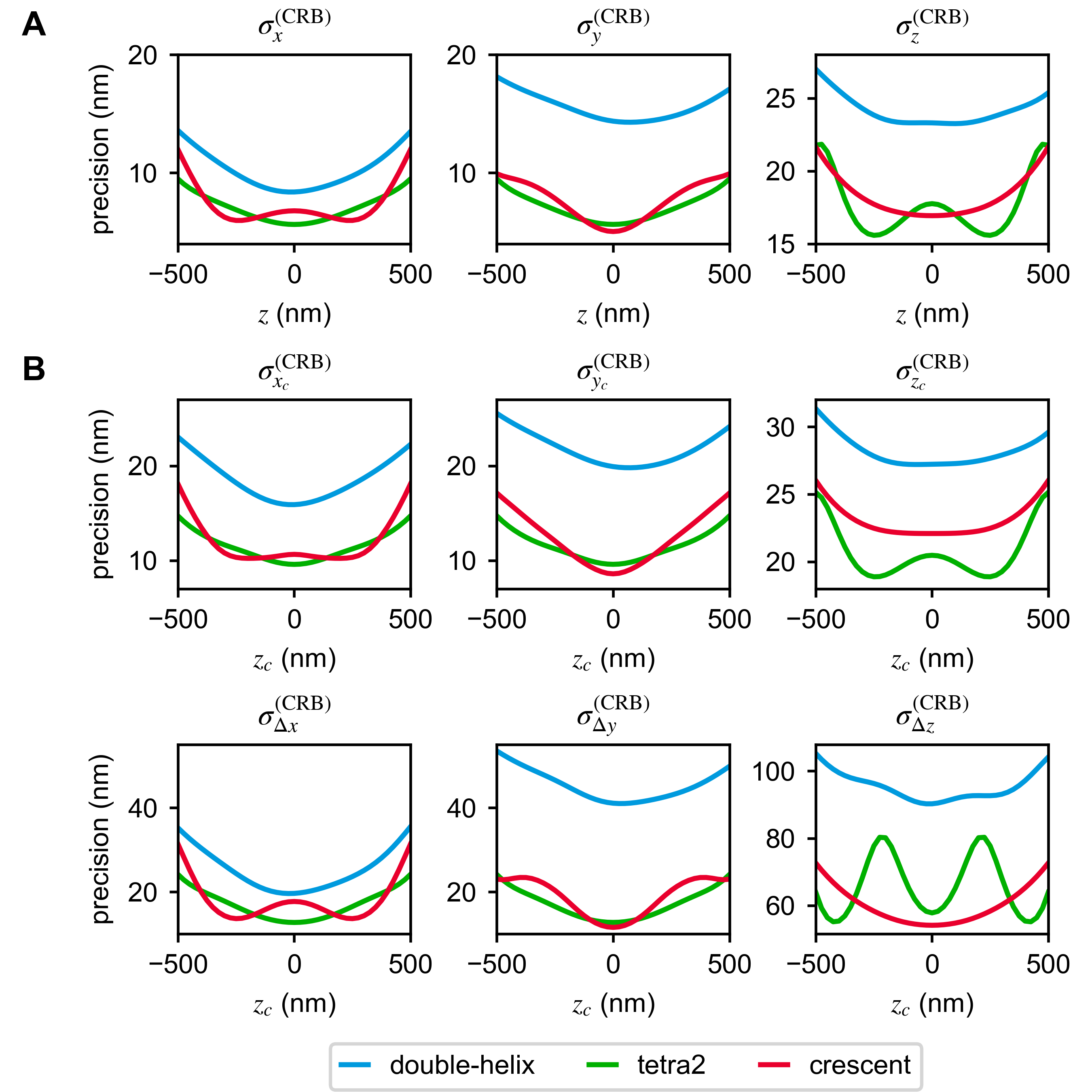}
\caption{Localization precision of the double-helix (cyan), tetra2 (green), and crescent (red) PSFs for (A)~imaging a single isolated emitter at axial position $z$ and (B)~two closely spaced emitters centered at axial position $z_c$. The precisions are calculated for emitters that are separated along the $x$-axis ($\sigma^\text{(CRB)}_{x_c}$ and $\sigma^\text{(CRB)}_{\Delta x}$), along the $y$-axis ($\sigma^\text{(CRB)}_{y_c}$ and $\sigma^\text{(CRB)}_{\Delta y}$), or along the $z$-axis ($\sigma^\text{(CRB)}_{z_c}$ and $\sigma^\text{(CRB)}_{\Delta z}$) by $200$~nm. All data are calculated using $N_\text{sig}=1000$~photons and $N_\text{bg}=10$~photons per pixel.}
\label{fig:locPrecisionCrescent}
\end{figure}

For the one-emitter case, we consider an emitter located at $(x_0, y_0, z_0)$, and we calculate $\sigma^\text{(CRB)}_x$, $\sigma^\text{(CRB)}_y$, and $\sigma^\text{(CRB)}_z$, the precisions of estimating $x_0$, $y_0$, and $z_0$ respectively from a noisy image (\cref{fig:locPrecisionCrescent}A). It should be noted that previously in this paper, Fisher information and CRBs were defined and used in the context of single-parameter estimation only, whereas now we calculate CRBs for multi-parameter estimation, using the $3\times 3$ Fisher information matrix for parameters $x_0$, $y_0$, and $z_0$ \cite{Kay1993}. This more accurately reflects the practical scenario where one needs to simultaneously estimate $x_0$, $y_0$ and $z_0$, and the CRBs may be affected by covariances between the parameters. The crescent PSF exhibits excellent localization precision ($\sigma^\text{(CRB)}_{x} = 7.3 \pm 1.7$~nm, $\sigma^\text{(CRB)}_{y} = 7.7 \pm 1.7$~nm, and $\sigma^\text{(CRB)}_{z} = 18.3 \pm 1.5$~nm, mean $\pm$ std over a depth range of 1~\textmu{}m using 1000 signal photons and 10 background photons per pixel).

For the two-emitter case (\cref{fig:locPrecisionCrescent}B), we consider emitter pairs separated along the $x$-, $y$-, and $z$-axes. For emitters at $(\pm100~\text{nm},0~\text{nm},z_c)$, we calculate $\sigma^\text{(CRB)}_{x_c}$, the precision of estimating the $x$-coordinate of the centroid $x_c$, as well as $\sigma^\text{(CRB)}_{\Delta x}$. For emitters at $(0~\text{nm},\pm100~\text{nm},z_c)$, we calculate $\sigma^\text{(CRB)}_{y_c}$ and $\sigma^\text{(CRB)}_{\Delta y}$. For axially separated emitters at $(x_c, y_c, z_c \pm 100~\text{nm})$, we calculate $\sigma^\text{(CRB)}_{z_c}$ and $\sigma^\text{(CRB)}_{\Delta z}$. Note that the plotted values of $\sigma^\text{(CRB)}_{\Delta x}$, $\sigma^\text{(CRB)}_{\Delta y}$, and $\sigma^\text{(CRB)}_{\Delta z}$ are exact values, unlike the approximations in \cref{eq:sigma_delta_x,eq:sigma_delta_y,eq:sigma_delta_z}. Furthermore, the centroid estimation CRBs are calculated using the Fisher information matrix for parameters $x_c$, $y_c$, and $z_c$, and the separation estimation CRBs are calculated using the Fisher information matrix for $\Delta x$, $\Delta y$, and $\Delta z$. The separation of $200$~nm is chosen specifically so that the two emitters would be difficult but not impossible to resolve. For reference, a standard microscope with the set of parameters used throughout this paper has an Abbe diffraction limit of $\lambda_0/2\mathrm{NA} = 196~\text{nm}$. Localization precision for two-emitter configurations with smaller and larger separations, as well as separations along multiple axes simultaneously, is addressed in Appendix \ref{subsec:CRBs_diff_separations}.

The crescent PSF performs well in resolving closely separated emitters in 3D without sacrificing performance in localizing single emitters. The crescent PSF outperforms tetra2 and double-helix in axial separation estimation precision ($\sigma^\text{(CRB)}_{\Delta z} = 60.2 \pm 5.7$~nm for crescent, compared to $\sigma^\text{(CRB)}_{\Delta z} = 65.5 \pm 8.4$~nm for tetra2 and $\sigma^\text{(CRB)}_{\Delta z} = 95.4 \pm 4.2$~nm for double-helix, mean~$\pm$~std over a depth range of 1~\textmu{}m using 1000 total signal photons and 10 background photons per pixel), and this performance is more uniform throughout the focal volume compared to tetra2. For estimating lateral separation, the crescent PSF performs similarly to tetra2 ($\sigma^\text{(CRB)}_{\Delta z} = 18.0 \pm 4.9$~nm for crescent, compared to $\sigma^\text{(CRB)}_{\Delta z} = 17.1 \pm 3.5$~nm for tetra2). Interestingly, the crescent PSF is also suitable for classic single-emitter localization, performing similarly to tetra2 and significantly better than double-helix, both of which were optimized for single-emitter localization ($\sigma^\text{(CRB)}_{x} = 7.3 \pm 1.7$~nm, $\sigma^\text{(CRB)}_{y} = 7.7 \pm 1.7$~nm, and $\sigma^\text{(CRB)}_{z} = 18.3 \pm 1.5$~nm for crescent; $\sigma^\text{(CRB)}_{x} = 7.1 \pm 1.2$~nm, $\sigma^\text{(CRB)}_{y} = 7.1 \pm 1.2$~nm, and $\sigma^\text{(CRB)}_{z} = 17.7 \pm 2.0$~nm for tetra2; $\sigma^\text{(CRB)}_{x} = 10.3 \pm 1.6$~nm, $\sigma^\text{(CRB)}_{y} = 15.6 \pm 1.1$~nm, and $\sigma^\text{(CRB)}_{z} = 24.2 \pm 1.0$~nm for double-helix).

\section{Accuracy of crescent PSF in distinguishing between one and two emitters}

To further quantify the crescent PSF's ability to resolve emitters with a small separation in the axial direction, we test its performance in distinguishing between noisy images of one emitter versus two emitters. We use a likelihood-ratio test to discriminate between one emitter at $(0, 0, z_c)$ versus two equally bright, incoherent emitters at $(0, 0, z_c \pm \Delta z/2)$ within a noisy image. To calculate the likelihood of each case, we begin by assuming a Poisson photon detection process, so the number of photons detected at a pixel at $(x_I, y_I)$ in image space is Poisson-distributed with mean $N_\text{sig}p(x_I, y_I) + N_\text{bg}$. Here, $N_\text{sig}$ is the total number of photons contained in the image, $p(x_I, y_I)$ is the probability distribution of photon detection in the image space, and $N_\text{bg}$ is the background flux in photons per pixel. For the one-emitter case, we use \cref{eq:prob-single-emitter} to find that the mean photon count at $(x_I, y_I)$, which we denote by $I_1(x_I, y_I)$, is given by
\begin{equation}
    I_1(x_I, y_I) = N_\text{sig}I(x_I, y_I; 0, 0, z_c) + N_\text{bg}.
\end{equation}
For the two-emitter case, we define $I_2(x_I, y_I)$ correspondingly, and it is given by
\begin{equation}
    I_2(x_I, y_I) = \frac{N_\text{sig}}{2}\left[I\left(x_I,y_I;0, 0, z_c + \frac{\Delta z}{2}\right) + I\left(x_I,y_I;0, 0, z_c - \frac{\Delta z}{2} \right)\right] + N_\text{bg}.
\end{equation}
Furthermore, denote the photon counts in the actual noisy image by $J(x_I,y_I)$. The likelihoods of the two scenarios are defined as the conditional probabilities of observing $J$ while assuming the number of emitters as ground truth \cite{Pitman1993}. Using the Poisson probability mass function, these are
\begin{linenomath}\begin{alignat}{2}
    \mathcal{L}(\text{1 emitter}) &= P(J | \text{1 emitter}) &&= \sum_{(x_I, y_I)} \frac{[I_1(x_I, y_I)]^{J(x_I,y_I)}}{\exp[I_1(x_I, y_I)][J(x_I,y_I)]!}
    \\
    \mathcal{L}(\text{2 emitters}) &= P(J | \text{2 emitters}) &&= \sum_{(x_I, y_I)} \frac{[I_2(x_I, y_I)]^{J(x_I,y_I)}}{\exp[I_2(x_I, y_I)][J(x_I,y_I)]!},
\end{alignat}\end{linenomath}
where the summations are taken over all pixels in the image. We conclude that there are two emitters if $\mathcal{L}(\text{2 emitters}) > \mathcal{L}(\text{1 emitter})$ and conclude one emitter otherwise.

We run the likelihood-ratio test on simulated noisy images of one and two emitters with the standard (no PM), double-helix, tetra2, and crescent PSFs, in three different conditions: $\Delta z = 100$~nm and $N_\text{sig}=10000$~photons, $\Delta z = 200$~nm and $N_\text{sig}=2000$~photons, and $\Delta z = 400$~nm and $N_\text{sig}=1000$~photons, all with $N_\text{bg}  =10$~photons per pixel and the same microscope parameters as in gradient descent. In each condition, we vary $z_c$ from $-500~\text{nm}$ to $500~\text{nm}$, and for each value of $z_c$, we run the test on $10000$ simulated images and calculate the fraction of runs for which it correctly predicts the number of emitters. Since single-molecule localization algorithms must perform joint detection and estimation simultaneously in practice, we note that this quantity is a best-case estimate of discrimination performance.

\begin{figure}[htb!]
\centering\includegraphics[width=5.25in]{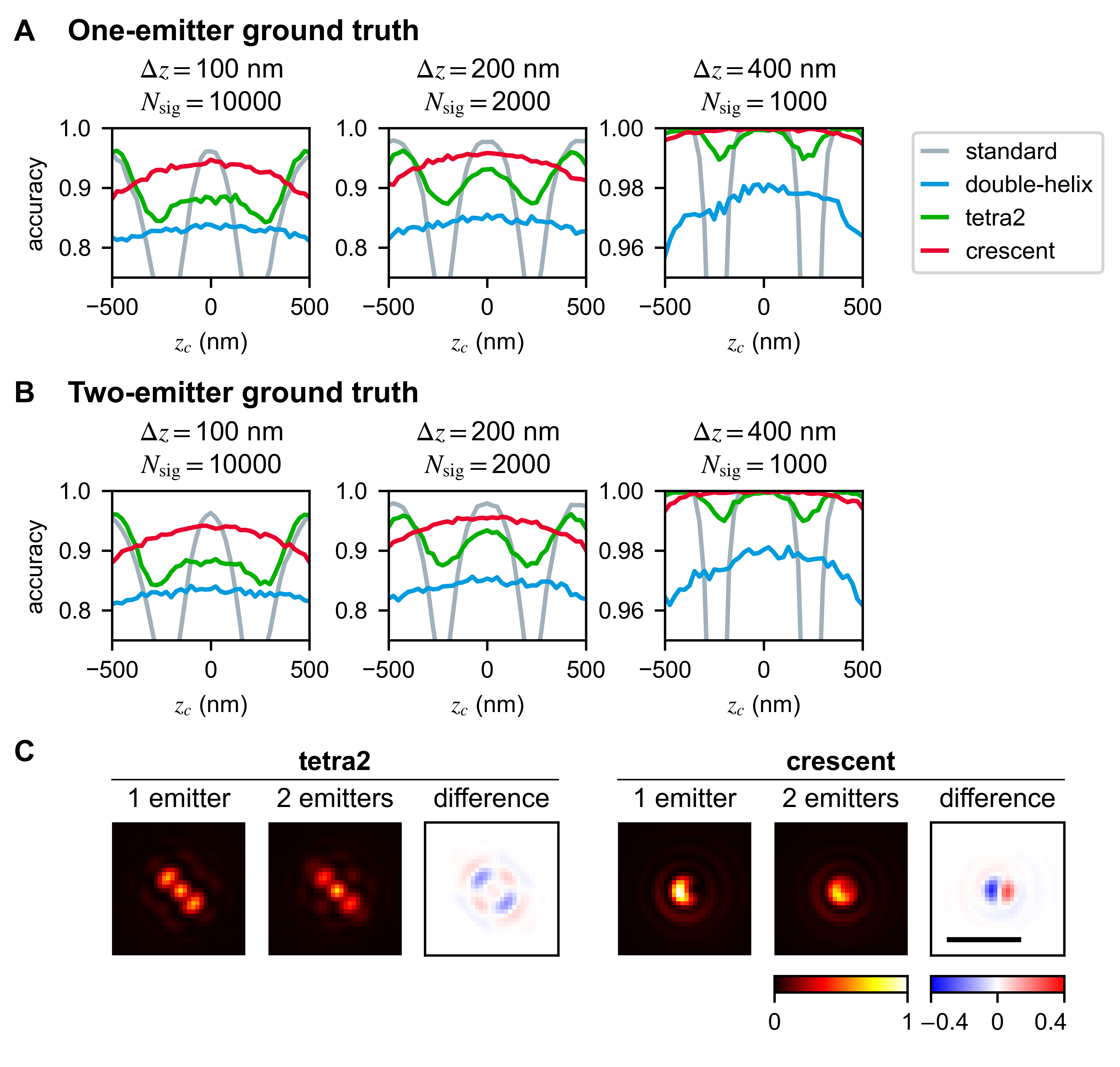}
\caption{Accuracy of distinguishing  between images containing one versus two emitters for the standard (gray), double-helix (cyan), tetra2 (green) and crescent (red) PSFs in three different conditions for (A)~one-emitter ground truth and (B)~two-emitter ground truth. Left: separation $\Delta z = 100$~nm and $N_\text{sig}=10000$~photons, middle: $\Delta z = 200$~nm and $N_\text{sig}=2000$~photons, and right: $\Delta z = 400$~nm and $N_\text{sig}=1000$~photons. (C)~Comparison of the images produced by one emitter ($z_0=200~\text{nm}$) and two emitters ($z_c=200~\text{nm}, \Delta z=400~\text{nm}$) with a constant signal level, using the tetra2 vs.\ crescent PSFs. To best visualize the advantage of the crescent PSF, $z_c$ and $\Delta z$ are chosen such that the crescent PSF performs significantly better than tetra2. The one-emitter image is subtracted from the two-emitter image to produce the difference image. For clarity, images are shown without Poisson noise and background noise. Color scales are in an arbitrary unit of intensity. The four 1-emitter and 2-emitter images share the same color scale, and the two difference images share the same color scale. Scale bar = 1~\textmu\text{m}.}
\label{fig:singleVsTwoEmitters}
\end{figure}

The crescent PSF is able to distinguish between the one- and two- emitter cases with an accuracy that is greater on average and more uniform as a function of $z_c$ compared to the other three PSFs (Figs.\ \ref{fig:singleVsTwoEmitters}A and \ref{fig:singleVsTwoEmitters}B). There are particular values of $z_c$ where the crescent is outperformed by tetra2, the next best-performing PSF, but the mean error rate of the crescent PSF is lower than that of tetra2 when averaged over the entire range of $z_c$. This improvement is consistent across all three conditions. For a two-emitter ground truth, for $\Delta z = \SI{100}{nm}$ and $N_\text{sig}=10000$, the mean error rate of the crescent PSF is 7.9\%, compared to 11.1\% for tetra2. For $\Delta z = \SI{200}{nm}$ and $N_\text{sig}=2000$, the mean error rates are 6.1\% for crescent and 8.2\% for tetra2. For $\Delta z = \SI{400}{nm}$ and $N_\text{sig}=1000$, the mean error rates are 0.15\% for crescent and 0.32\% for tetra2. The standard and double-helix PSFs perform significantly worse than both crescent and tetra2. These values are similar for a one-emitter ground truth.

One qualitative observation that may explain the crescent PSF's advantage in resolving axially separated emitters is that with the crescent PSF, the images of the one- and two-emitter configurations differ more greatly than with tetra2 (\cref{fig:singleVsTwoEmitters}C). This contrast is due to the crescent PSF's compact shape, which allows it to concentrate light into a relatively small area, resulting in higher photon counts that enable more accurate discrimination between overlapping emitters versus Poisson shot noise.

\section{Conclusion}
In this paper, we studied the ability of engineered PSFs to resolve closely spaced emitters in 3D. Using simple mathematical reasoning, we developed a heuristic for optimizing the precision of estimating the separation distance between two emitters. Gradient descent optimization of our cost function revealed a new type of PSF, which we call the crescent PSF, that rotates as a function of emitter depth but has a simpler PM design compared to existing corkscrew-like PSFs \cite{Lew2011,Prasad2013,Nehme2020,Nehme2021}. Quantifying estimation precision in terms of the Cram\'er-Rao bound, we showed that the crescent PSF performs 8\% better than the best Tetrapod PSF in estimating the separation between emitters along the axial direction, and its performance is also more uniform throughout the focal volume. The crescent PSF also performs well in single-emitter localization, similar to the Tetrapod PSF. Lastly, we showed that with a simple likelihood-ratio test, the crescent PSF allows for distinguishing between the noisy images produced by one versus two emitters with a 29-53\% lower error rate, averaged throughout the focal volume, than with the Tetrapod PSF.

Our study primarily relies on numerical methods to explore the design space and assess performance. Starting the optimization from existing engineered PSFs serves to direct the algorithm to possible minima near a high-performing PSF (exploitation), while starting from random conditions ensures that many areas of the parameter space are surveyed (exploration). Of course, we cannot disprove the existence of additional sharp local minima in the cost function. While the crescent PSF outperforms existing PSFs engineered for 3D single-emitter localization in many cases, our data do not prove that is optimal for all imaging tasks. Further work remains to design optimal PSFs for imaging within thick specimens, especially in the presence of optical aberrations \cite{Ghosh2015,Mlodzianoski2018,Xu2020}.

Our results demonstrate that localizing single emitters, resolving closely spaced emitters in the lateral direction and resolving them in the axial direction are three very different problems, and the PSFs that are optimal for one task are not necessarily optimal for another. As a result, we believe that simulating carefully designed emitter configurations, as we did in Section 4, can help experimentalists choose and/or design PSFs that are optimal for their particular application.

It is notable that optimizing the cost function $C_2$ (\cref{eq:C2}), a second-order approximation of the CRB for estimating the separation between two emitters, resulted consistently in two types of PSFs: one rotating single-spot PSF and another translating double-spot PSF. These PSFs are created by PMs with remarkable symmetry and simple phase profiles. We speculate that these PMs, which both separate Fourier space into a central disk and surrounding ring, are related to optimal axial localization of single emitters via interferometry \cite{Backlund2018}. That is, both PMs contain a central disk whose radius is \textasciitilde{}70\% that of the aperture, and this size is very similar to that of the annular mirror ($r/r_A=0.63$) within the optimal interferometer in Ref.~\cite{Backlund2018}. It is likely that these rings in Fourier space are ideal locations for discriminating wavefront curvature as an emitter becomes defocused. Further studies on the relationship between detection, estimation, and resolution of multiple point emitters, and the role of annular-style PMs, will be insightful for pushing 3D super-resolution imaging to its ultimate limits. 


\clearpage
\section{Appendix}

\subsection{Errors in approximate expressions for $\sigma^\text{(CRB)}_{\Delta x}$, $\sigma^\text{(CRB)}_{\Delta y}$, and $\sigma^\text{(CRB)}_{\Delta z}$}\label{subsec:CRBapprox}

\begin{figure}[ht!]
\centering\includegraphics[width=5.25in]{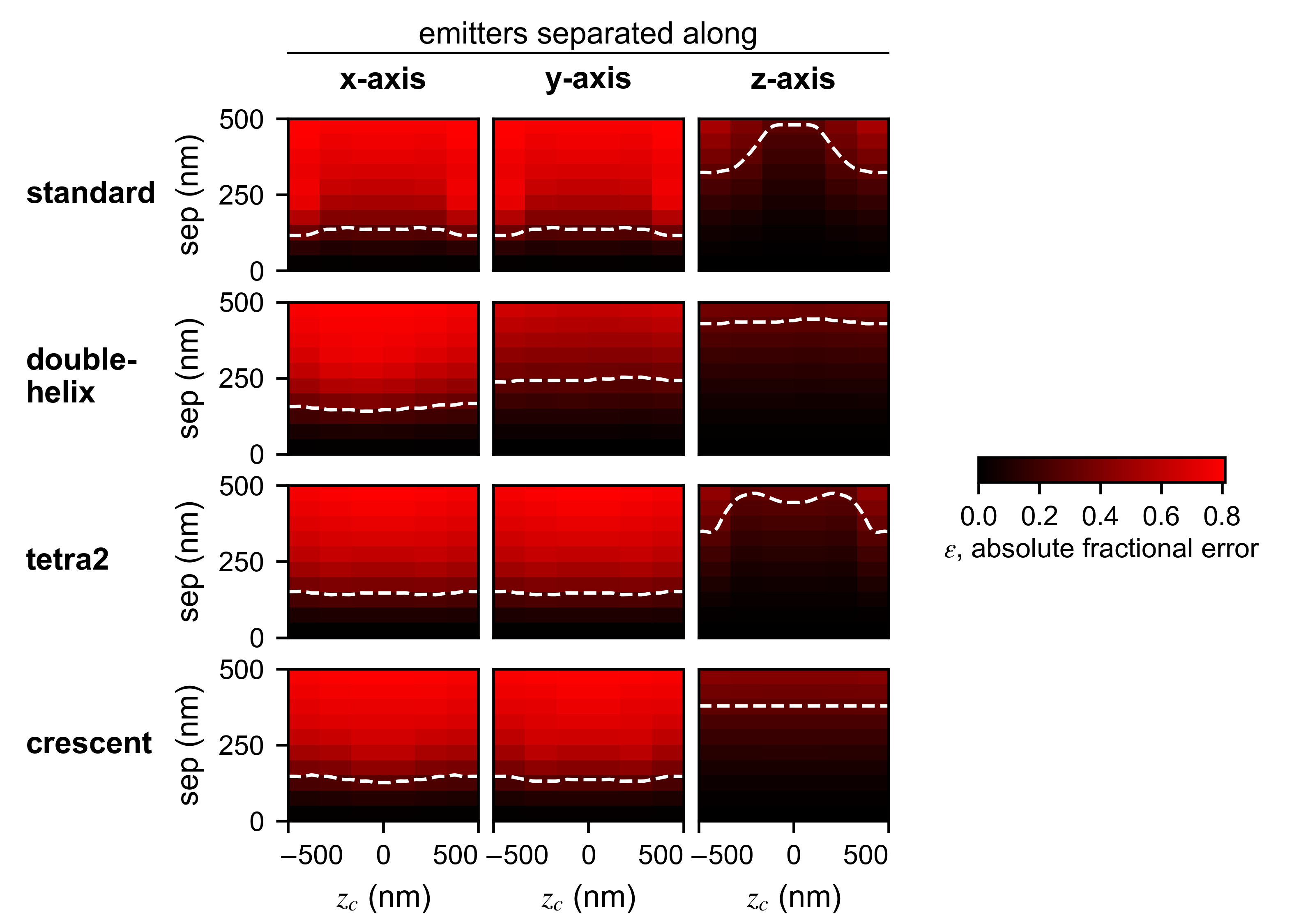}
\caption{Absolute fractional errors (\cref{eq:abs_frac_error}) in approximate expressions for $\sigma^\text{(CRB)}_{\Delta x}$, $\sigma^\text{(CRB)}_{\Delta y}$, and $\sigma^\text{(CRB)}_{\Delta z}$ (\cref{eq:sigma_delta_x,eq:sigma_delta_y,eq:sigma_delta_z} respectively) for various PSFs. Errors are plotted as a function of axial centroid $z_c$ and appropriate emitter separation $\Delta x$, $\Delta y$, or $\Delta z$. The color scale represents the absolute value of the fractional error of the expression, relative to its true value. The dashed white lines are contour lines at a fixed error of $\pm 0.3=\pm 30\%$. All data are calculated using $N_\text{sig}=1000$~photons and $N_\text{bg}=10$~photons per pixel.}
\label{fig:CRBapprox}
\end{figure}

We evaluate the error in our approximate expressions for $\sigma^\text{(CRB)}_{\Delta x}$, $\sigma^\text{(CRB)}_{\Delta y}$, and $\sigma^\text{(CRB)}_{\Delta z}$, i.e., \cref{eq:sigma_delta_x,eq:sigma_delta_y,eq:sigma_delta_z}, respectively, for various PSFs. For a given PSF, separation distance, and $z_c$, if the approximate value of a precision bound is $\sigma_\text{approx}$ and its actual value is $\sigma_\text{actual}$, we calculate the absolute fractional error, defined as
\begin{equation}\label{eq:abs_frac_error}
    \varepsilon = \frac{|\sigma_\text{approx}-{\sigma_\text{actual}|}}{\sigma_\text{actual}}.
\end{equation}
The results are shown for the standard (no PM), double-helix, tetra2, and crescent PSFs in \cref{fig:CRBapprox}.

Although the approximations begin to deviate significantly from their true values for separations of a few hundred nanometers, the approximations are correct within $30\%$ for lateral separations of up to 100-200 nm and axial separations of up to 300-500 nm. Since we are concerned with resolving closely spaced emitters (whose separation is perhaps on the same order as the Abbe diffraction limit, $\lambda_0/2\mathrm{NA} = 196~\text{nm}$), these approximations are appropriate to use, thus greatly reducing the computational complexity of the optimization task in \cref{subsec:optimization}. While these approximate CRBs are convenient for optimization, we use exact expressions for the CRB when evaluating PSF performance.

\subsection{New Tetrapod-like PSF from gradient descent}\label{subsec:tetraLikePMs}

\begin{figure}[ht!]
\centering\includegraphics[width=5.25in]{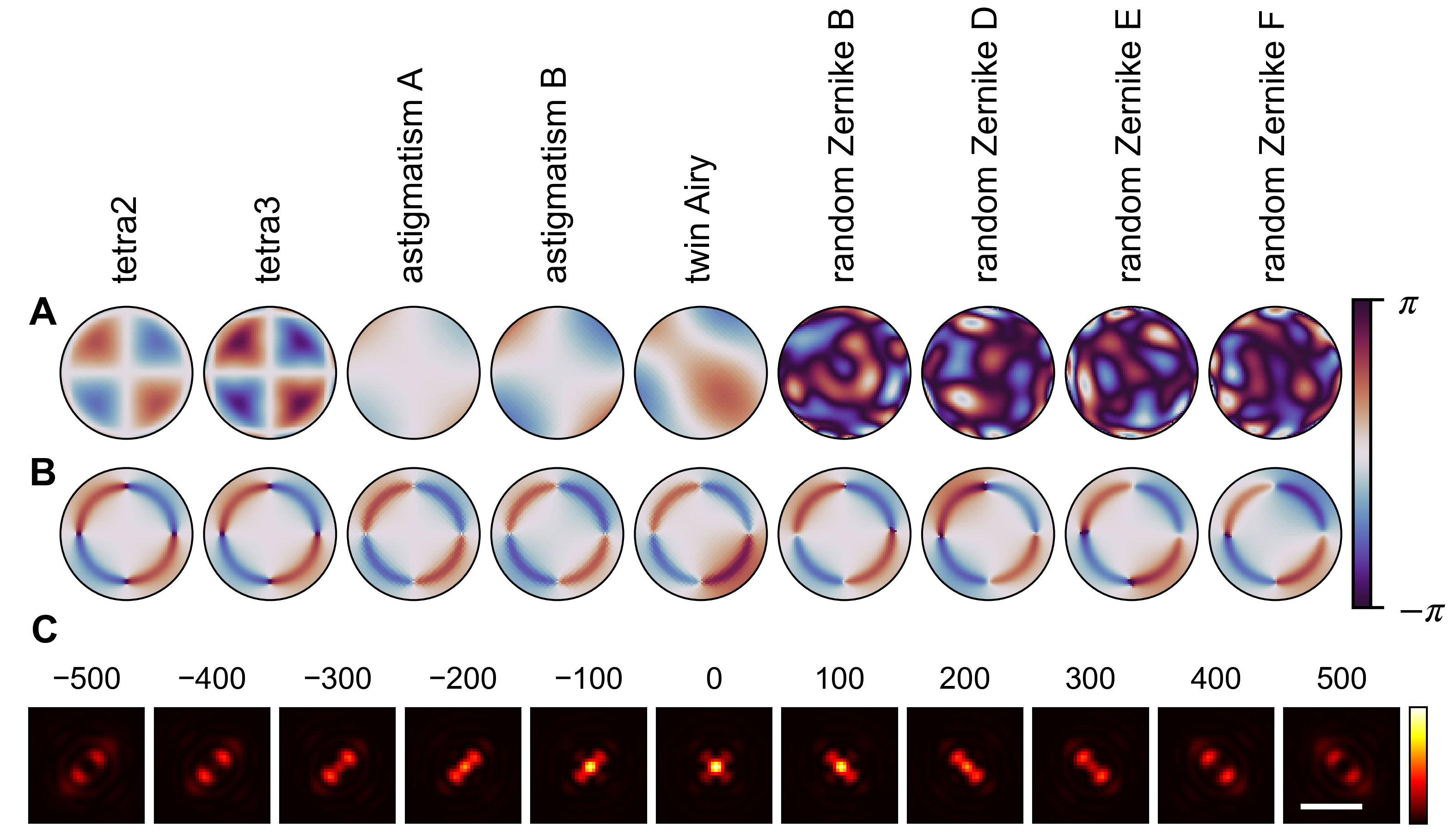}
\caption{Optimality and stability of the new Tetrapod-like PSF from gradient descent. (A)~The various initial PMs that converge to a new Tetrapod-like PSF. (B)~The final PMs, standardized as described in Appendix \ref{subsec:normalizingPMs}. (C)~The Tetrapod-like PSF generated by the final PM from the tetra2 initial condition. The color scale represents intensity. Scale bar = 1~\textmu{}m. The PSFs generated by the other final PMs are virtually indistinguishable and are therefore not shown.}
\label{fig:tetraLikePMs}
\end{figure}

\begin{figure}[ht!]
\centering\includegraphics[width=4.5in]{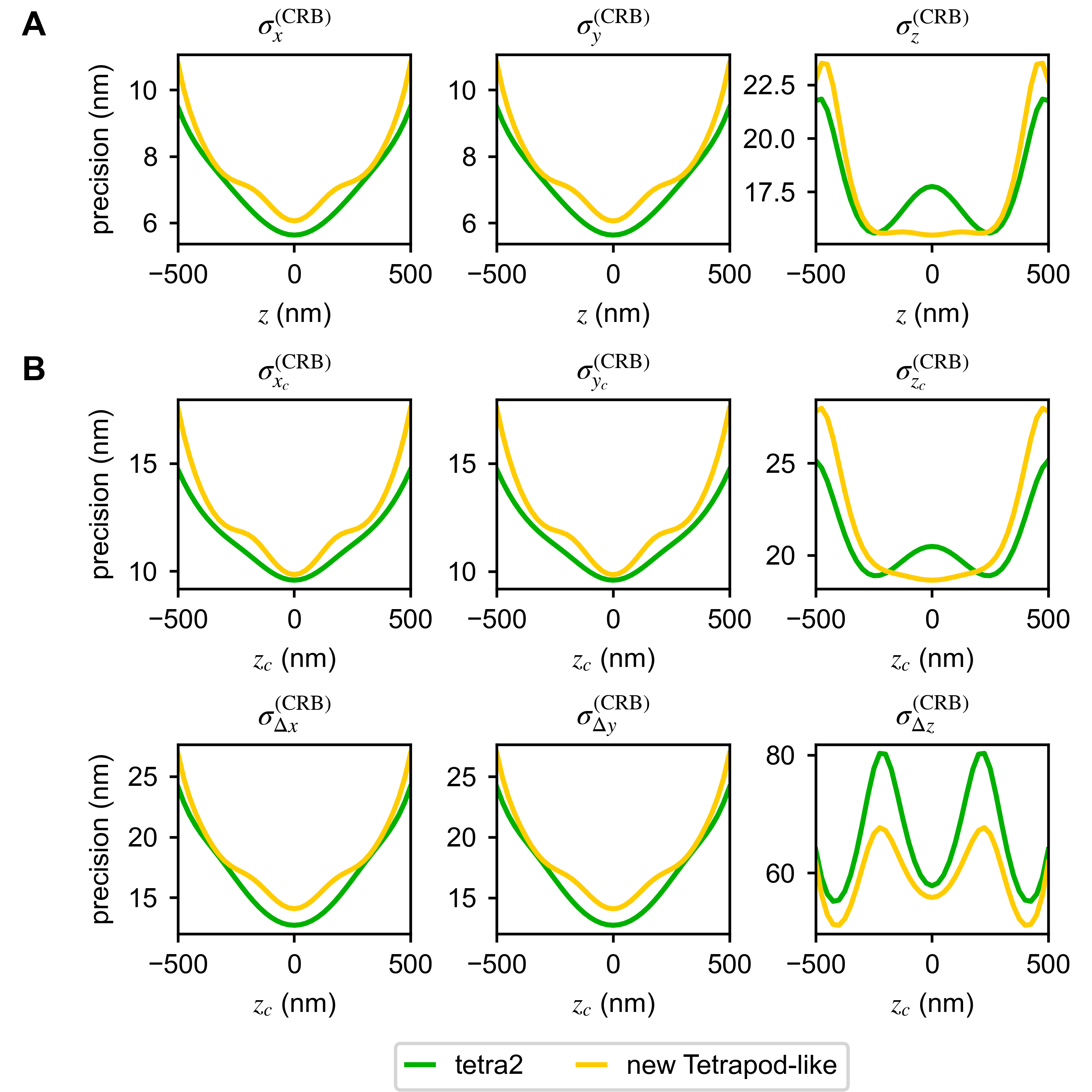}
\caption{Localization precision of the tetra2 (green) and new Tetrapod-like (gold, \cref{fig:tetraLikePMs}C) PSFs for imaging (A)~a single isolated emitter at axial position $z$ and (B)~two closely spaced emitters centered at position $z_c$. The precisions are calculated for emitters that are separated along the $x$-axis ($\sigma^\text{(CRB)}_{x_c}$ and $\sigma^\text{(CRB)}_{\Delta x}$), along the $y$-axis ($\sigma^\text{(CRB)}_{y_c}$ and $\sigma^\text{(CRB)}_{\Delta y}$), or along the $z$-axis ($\sigma^\text{(CRB)}_{z_c}$ and $\sigma^\text{(CRB)}_{\Delta z}$) by $200$~nm. All data are calculated using $N_\text{sig}=1000$~photons and $N_\text{bg}=10$~photons per pixel.}
\label{fig:tetra2_vs_tetraLike}
\end{figure}

The initial PMs and final PMs that converge to a two-spot translating PSF are shown in \cref{fig:tetraLikePMs}. The behavior of the PSF as a function of emitter depth is similar to those in the existing Tetrapod family. The performance of our new Tetrapod-like PSF produced by gradient descent is very similar to that of the existing tetra2 PSF (\cref{fig:tetra2_vs_tetraLike}), so we do not report on it further.

\clearpage
\subsection{Convergence of cost function in gradient descent}\label{subsec:costFxnDescent}

\begin{figure}[ht!]
\centering\includegraphics[width=3.75in]{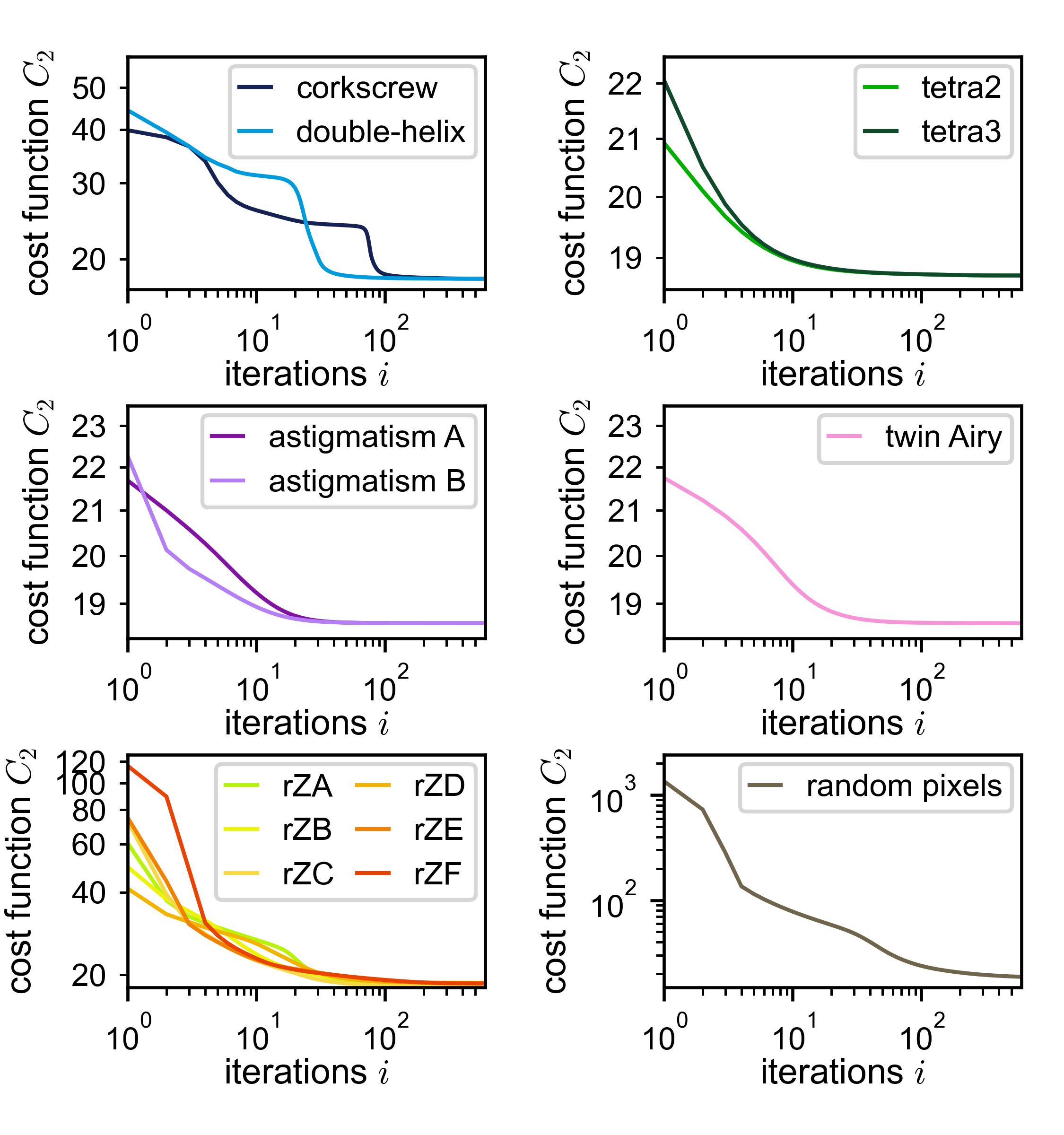}
\caption{Cost function $C_2$ (\cref{eq:C2}) versus number of elapsed iterations $i$ for all runs of gradient descent (\cref{alg:grad_descent}). Legend labels indicate the initial PM of the run; rZ = random Zernike.}
\label{fig:costFxnDescent}
\end{figure}

Each run of gradient descent is run for a total of 600 iterations, which is sufficient for the cost function $C_2$ (\cref{eq:C2}) to converge upon a local minimum in all cases (\cref{fig:costFxnDescent}).

\clearpage
\subsection{Standardization of the final phase masks from gradient descent}\label{subsec:normalizingPMs}
\begin{figure}[ht!]
\centering\includegraphics[width=4.75in]{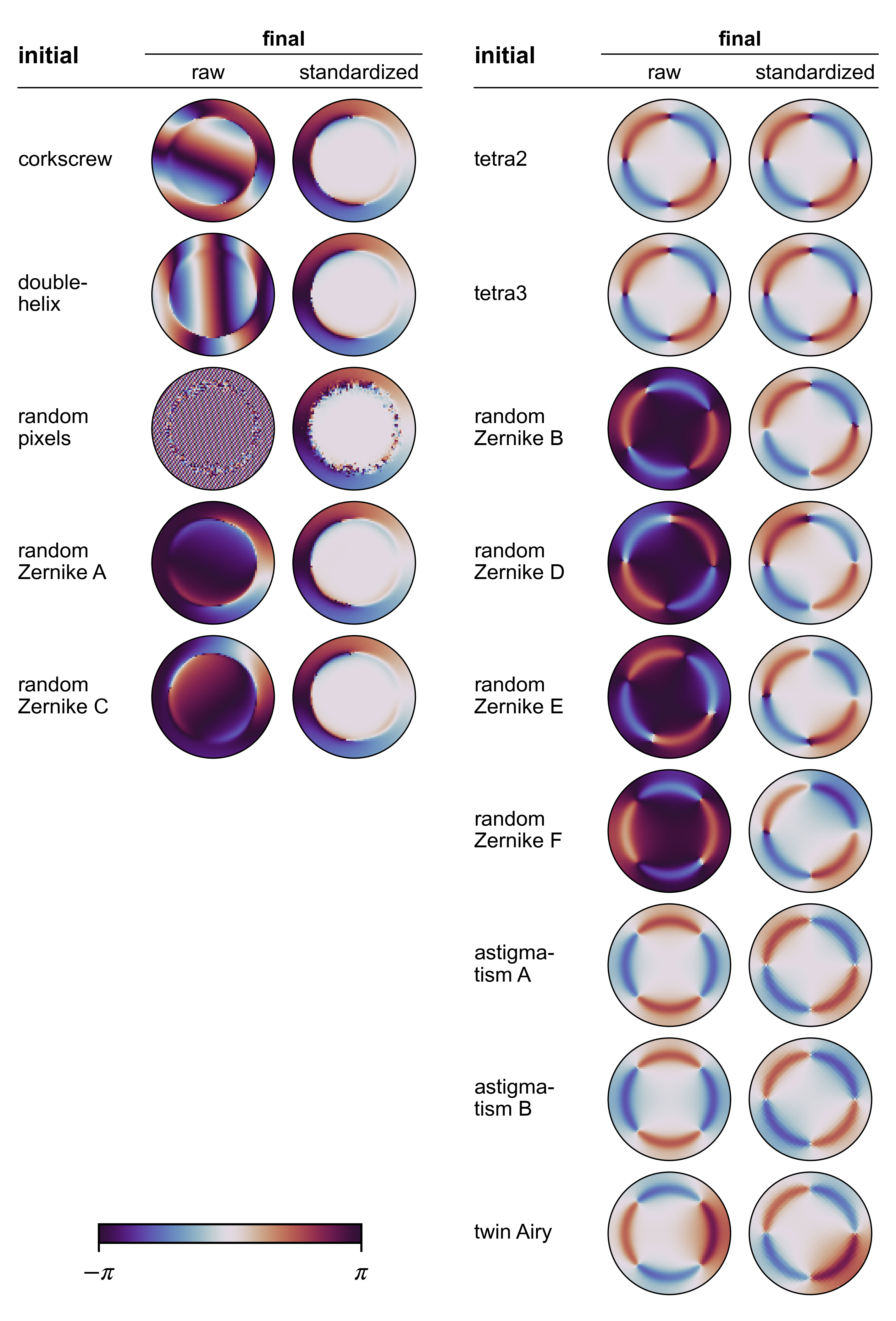}
\caption{Standardization of final PMs from \cref{alg:grad_descent}. The colorbar represents phase in radians.}
\label{fig:normalizingPMs}
\end{figure}
\noindent The gradient descent algorithm imparts no constraints on the location of the PSF in the image nor on the orientation of the PSF. Therefore, the final PSF outputted by the algorithm is not necessarily centered in the image and can have an arbitrary rotation. In order to ensure that similar PSFs appear visually similar to the reader, we standardize the final PSFs from gradient descent by translating them to be centered in the image (by adding a linear function to the PM) and rotating them to match a common orientation (by rotating the PM by the desired amount) (\cref{fig:normalizingPMs}). The appropriate linear functions and rotation angles are chosen by hand.

\subsection{Creating an idealized phase mask for the crescent PSF}\label{subsec:idealizedCrescentPM}

\begin{figure}[ht!]
\centering\includegraphics[width=4.75in]{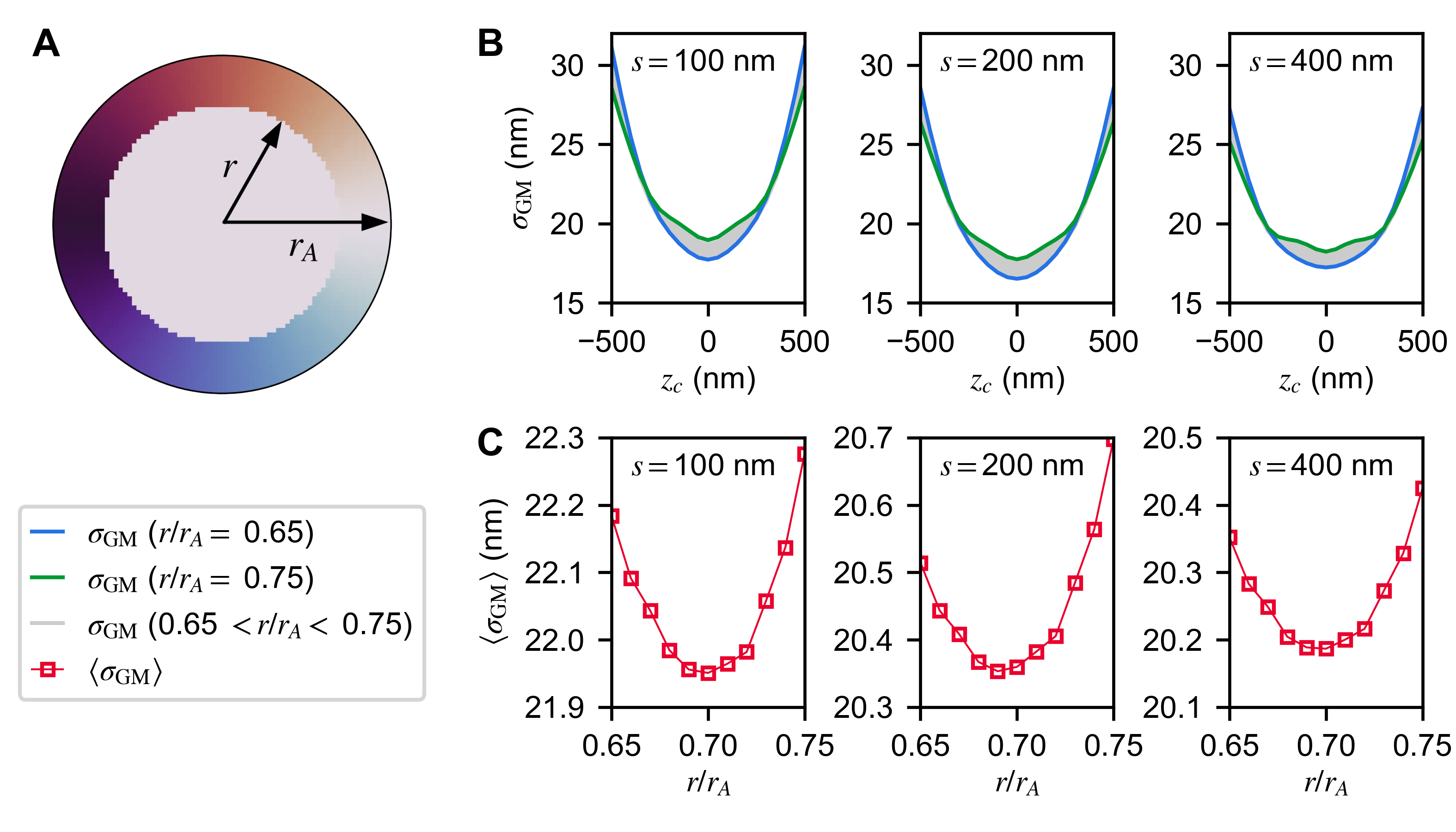}
\caption{Creating an idealized PM for the crescent PSF. (A)~The idealized PM consists of an outer ring, where the phase shift is equal to the polar angle, and an inner circle, where the phase shift is zero. We denote radius of the boundary between the two regions as $r$ and the radius of the aperture as $r_A$. (B)~Mean precision $\sigma_\text{GM}$ vs.\ centroid location $z_c$ for $0.65 \leq r/r_A \leq 0.75$. The blue curve represents $r/r_A = 0.65$, the green curve represents $r/r_A = 0.75$, and curves for intermediate values of $r/r_A$ lie in the gray region. (C)~Mean precision $\langle \sigma_\text{GM} \rangle$ vs.\ $r/r_A$. All data are calculated using $N_\text{sig}=1000$~photons and $N_\text{bg}=10$~photons per pixel.}
\label{fig:idealizedCrescentPM}
\end{figure}

The five final PMs from gradient descent with ring-shaped ramps around circular regions of constant zero phase are essentially identical except for minor imperfections at the boundary between the two regions (\cref{fig:crescentPMs}B). To create an idealized version of the crescent PM, we remove these imperfections, and we vary the radius of the boundary between the two regions to find the value that results in the best performance (\cref{fig:idealizedCrescentPM}A). Specifically, we would like the crescent PSF to simultaneously perform well in estimating $x_c$, $y_c$, $z_c$, $\Delta x$, $\Delta y$, and $\Delta z$ for configurations of two closely spaced emitters. Define the quantity
\begin{equation}
    \sigma_\text{GM} = \sqrt{\sigma^\text{(CRB)}_{x_c}\sigma^\text{(CRB)}_{y_c}\sigma^\text{(CRB)}_{z_c}\sigma^\text{(CRB)}_{\Delta x}\sigma^\text{(CRB)}_{\Delta y}\sigma^\text{(CRB)}_{\Delta z}},
\end{equation}
where $\sigma^\text{(CRB)}_{\Delta x}$ is calculated for emitters at $(\pm s, 0, z_c)$, $\sigma^\text{(CRB)}_{\Delta y}$ is calculated for emitters at $(0, \pm s, z_c)$, and $\sigma^\text{(CRB)}_{\Delta z}$ is calculated for emitters at $(0, 0, z_c \pm s)$. Then $\sigma_\text{GM}$ serves as a general measure of performance of the PSF for fixed values of $s$ and $z_c$ (\cref{fig:idealizedCrescentPM}B). Next, define $\langle \sigma_\text{GM}\rangle$ to be the average value of $\sigma_\text{GM}$ for $z_c \in \{-500\text{ nm}, 500\text{ nm}\}$. Based on the values of $\langle \sigma_\text{GM} \rangle$ as a function of PM's inner radius for $s=$ 100, 200, and 400 nm (\cref{fig:idealizedCrescentPM}C), we choose the final design of the crescent PM to have an inner radius equal to 0.69 times the radius of the aperture. This final PM and its corresponding PSF are shown in \cref{fig:Schematic}B and \cref{fig:Schematic}C, respectively.

\subsection{Precision of the crescent PSF for two-emitter localization with varying separation distances and directions}\label{subsec:CRBs_diff_separations}
\begin{figure}[ht!]
\centering\includegraphics[width=4.5in]{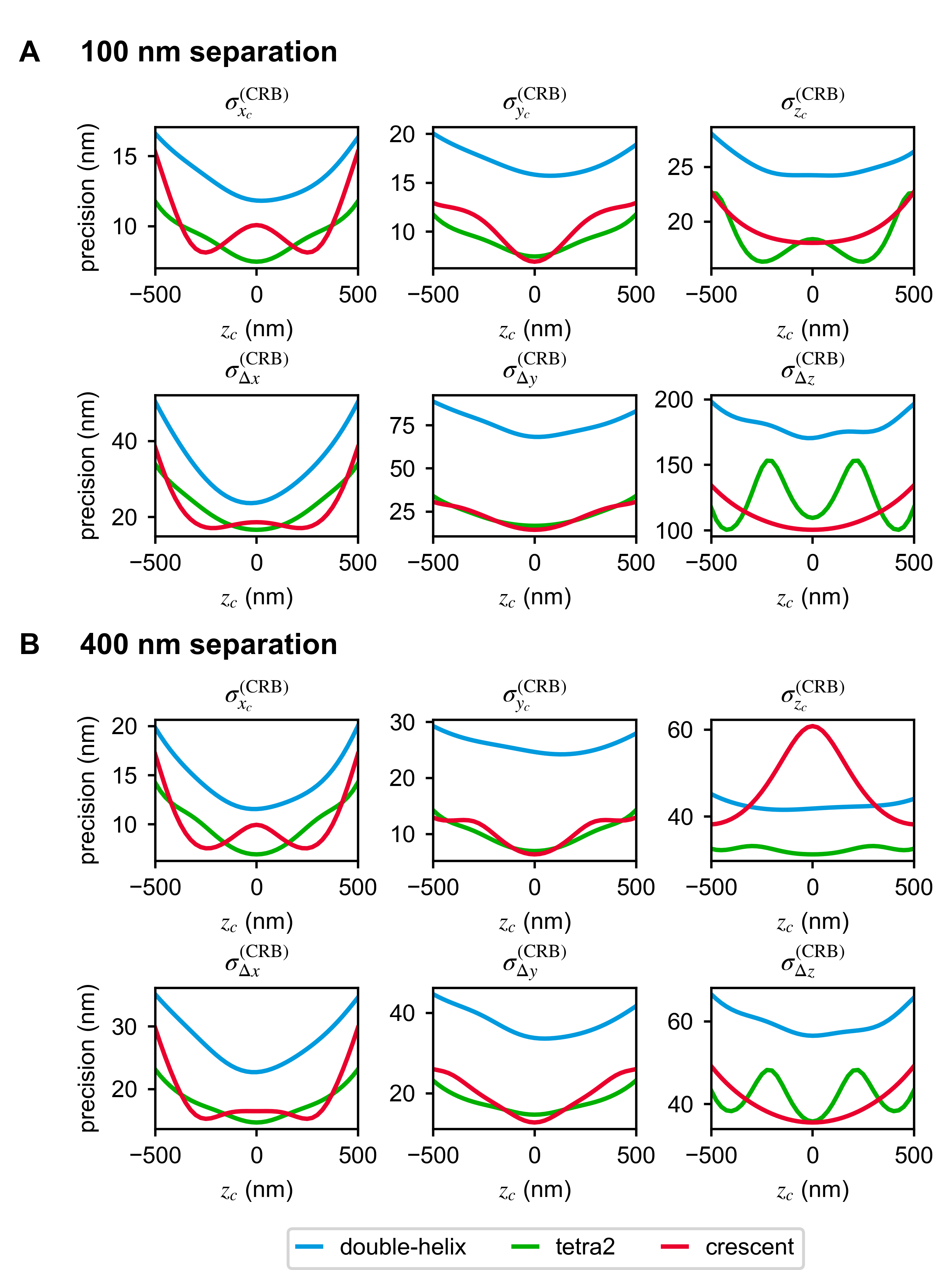}
\caption{Localization precision of the double-helix (cyan), tetra2 (green), and crescent (red) PSFs for imaging two closely spaced emitters centered at axial position $z_c$. The precisions are calculated for emitters that are separated along the $x$-axis ($\sigma^\text{(CRB)}_{x_c}$ and $\sigma^\text{(CRB)}_{\Delta x}$), along the $y$-axis ($\sigma^\text{(CRB)}_{y_c}$ and $\sigma^\text{(CRB)}_{\Delta y}$), or along the $z$-axis ($\sigma^\text{(CRB)}_{z_c}$ and $\sigma^\text{(CRB)}_{\Delta z}$) by (A)~100~nm or (B)~400~nm. All data are calculated using $N_\text{sig}=1000$~photons, and $N_\text{bg}=10$~photons per pixel.
}
\label{fig:CRBs_diff_sep}
\end{figure}
\begin{figure}[ht!]
\centering\includegraphics[width=4.5in]{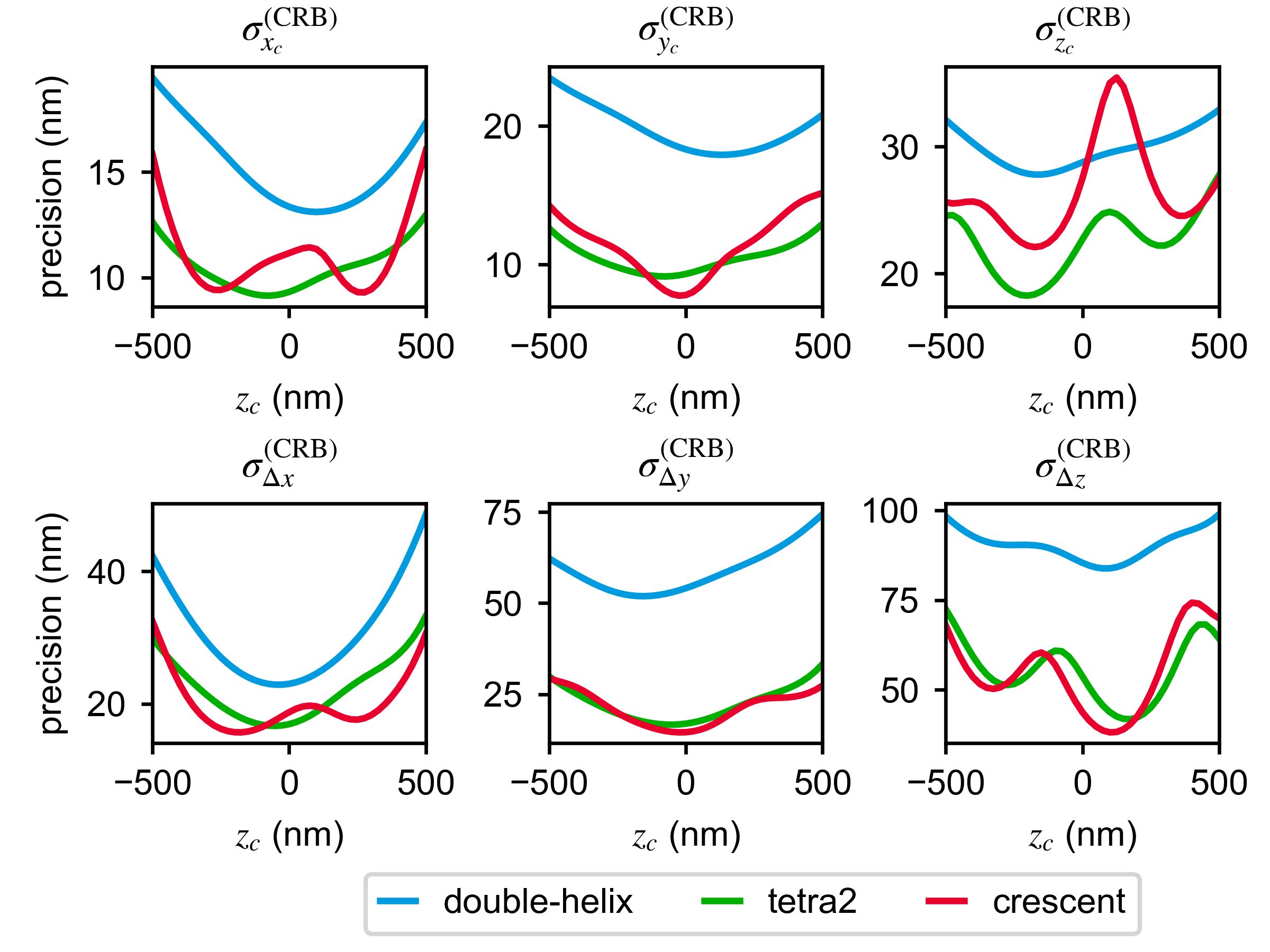}
\caption{Localization precision of the double-helix (cyan), tetra2 (green), and crescent (red) PSFs for imaging two closely spaced emitters centered at axial position $z_c$. The precisions are calculated for emitter 1 at $x_1=y_1=z_1 = (\SI{200}{nm})/\sqrt{3}$ and emitter 2 at $x_2=y_2=z_2 = -(\SI{200}{nm})/\sqrt{3}$. All data are calculated using $N_\text{sig}=1000$~photons and $N_\text{bg}=10$~photons per pixel.
}
\label{fig:CRBs_diagonal_sep}
\end{figure}

\noindent \cref{fig:locPrecisionCrescent} only shows the two-emitter Cram\'{e}r-Rao bounds for emitter separations of 200~nm along the axis corresponding to the quantity being estimated. Here, we consider different values of this separation distance, as well as simultaneous separations along multiple axes. \cref{fig:CRBs_diff_sep} shows the two-emitter localization precisions for emitter separations of 100~nm and 400~nm. \cref{fig:CRBs_diagonal_sep} shows the localization precisions for two emitters separated by 200~nm with an equal component along each axis, i.e.\ one emitter at $x_1=y_1=z_1 = (\SI{200}{nm})/\sqrt{3}$ and the other at $x_2=y_2=z_2 = -(\SI{200}{nm})/\sqrt{3}$. In both cases, the relative performances of the three PSFs are similar to those in \cref{fig:locPrecisionCrescent}, with a few exceptions.

\begin{backmatter}
\bmsection{Funding} Research reported in this publication was supported by the National Science Foundation under grant number ECCS-1653777 to M.D.L.
\bmsection{Acknowledgments} We thank W.E.\ Moerner and Yoav Shechtman for providing the Tetrapod phase masks. J.M.J.\ acknowledges support from the Caltech Summer Undergraduate Research Fellowship program. Special thanks to Yiyang Chen for a critical reading of the manuscript.
\bmsection{Disclosures} The authors declare no conflicts of interest.
\bmsection{Data Availability} Data underlying the results presented in this paper are available in Ref.\ \cite{OSFrepository} and may also be obtained from the authors upon reasonable request.
\end{backmatter}

\bibliography{references}

\end{document}